\documentclass[12pt,preprint]{aastex}

\usepackage{latexsym}
\usepackage{graphicx}
\usepackage{epsfig}
\usepackage{psfig}
\usepackage{natbib}
\usepackage{amsmath}
\usepackage{amssymb}

\def\mum{${\rm \mu m}$}
\def\HII{\ion{H}{2}~} 
\def\jpc{Jour.~Phys.~Chem.}
\def\jpca{Jour.~Phys.~Chem.~A}

\shorttitle{The profiles of the 3 to 12 $\mu$m PAH features.}
\shortauthors{van Diedenhoven et al.}

\begin{document}

\title{The profiles of the 3 to 12 $\mu$m PAH features\altaffilmark{1}.}

\author{B. van Diedenhoven\altaffilmark{2},
E. Peeters\altaffilmark{3,2,4}, C. Van Kerckhoven\altaffilmark{5},
S. Hony\altaffilmark{6}, D.M. Hudgins\altaffilmark{4},\\
L.J. Allamandola\altaffilmark{4}, A.G.G.M. Tielens\altaffilmark{3,2}}

\altaffiltext{1}{Based on observations with ISO, an ESA project with
    instruments funded by ESA Member States (especially the PI
    countries: France, Germany, the Netherlands and the United
    Kingdom) and with the participation of ISAS and NASA.}
    \altaffiltext{2}{Kapteyn Institute, P.O. Box 800, 9700 AV
    Groningen, The Netherlands}
    \altaffiltext{3}{SRON~National~Institute~for~Space~Research,
    P.O. Box 800, 9700 AV Groningen, The Netherlands}
    \altaffiltext{4}{NASA-Ames Research Center, Mail Stop 245-6,
    Moffett Field, CA 94035, USA; epeeters@mail.arc.nasa.gov}
    \altaffiltext{5}{Instituut voor Sterrenkunde, K.U.Leuven,
    Celestijnlaan 200B, 3100 Heverlee, Belgium}
    \altaffiltext{6}{RSSD-ESA/ESTEC, PO Box 299, 2200 AG Noordwijk,
    The Netherlands}

\begin{abstract}

We present spectra of the 3.3 $\mu$m and 11.2 $\mu$m PAH features of a
large number of stellar sources, planetary nebulae, reflection
nebulae, \HII regions and galaxies, obtained with ISO-SWS. Clear
variations are present in the profiles of these features. Most of the
sources show a symmetric 3.3 $\mu$m feature peaking at $\sim$3.290
$\mu$m, while only very few show an asymmetric 3.3 $\mu$m feature
peaking at a slightly longer wavelength. The profiles of the 11.2
\mum\, feature are distinctly asymmetric. The majority of the sources
has a 11.2 \mum\, feature peaking between 11.20 and 11.24 \mum, with a
very steep blue rise and a low tail-to-top ratio. A few sources show a
11.2 \mum\, feature with a peak position of $\sim$11.25 \mum, a less
steep blue rise and a high tail-to-top ratio. The sources are
classified independently based on the 3.3 and 11.2 $\mu$m feature
profiles and peak positions.  Correlations between these classes and
those based on the 6--9 $\mu$m features \citep{Peeters:prof6:02} are
found. In particular, sources with the most common profiles in the
6--9 $\mu$m region also show the most common 3.3 and 11.2 $\mu$m
feature profiles. However, the uncommon profiles do not correlate with
each other. Also, these classifications depend on the type of
object. In general, \HII regions, non-isolated Herbig AeBe stars and
YSO's show the same profiles for all 3-12 $\mu$m features. Many
planetary nebulae and Post-AGB stars show uncommon feature
profiles. The 3 galaxies in our sample show the same profiles as the
\HII regions for all but the 11.2 \mum\, feature, being similar to
that of evolved stars. The observed pronounced contrast in the
spectral variations for the CH modes (3.3 and 11.2 \mum\, bands)
versus the CC modes (6.2, 7.7 and 8.6 \mum\, bands) is striking : the
peak wavelengths of the features attributed to CC modes vary by
$\sim$15--80 cm$^{-1}$, while for the CH modes the variations are
$\sim$4--6.5 cm$^{-1}$. We summarize existing laboratory data and
theoretical calculations of the modes emitting in the 3--12 \mum\,
region of PAH molecules and complexes.  In contrast to the 6.2 and 7.7
\mum\, components which are attributed to PAH cations, the 3.3 \mum\,
feature appears to originate in neutral and/or negatively charged
PAHs. We attribute the variations in peak position and profile of
these IR emission features to the composition of the PAH family. The
variations in FWHM of the 3.3 \mum\, feature remains an enigma while
those of the 11.2 \mum\, can be explained by anharmonicity and
molecular structure. The possible origin of the observed contrast in
profile variations between the CH modes and the CC modes is
highlighted.
\end{abstract}


\section{Introduction}
\label{chvscc_INTRO}
The infrared spectra of a wide variety of sources are dominated by
strong emission features at 3.3, 6.2, 7.7, 8.6 and 11.2 $\mu$m (3040,
1615, 1310, 1160, 890 cm$^{-1}$), commonly called the unidentified
infrared (UIR) emission features \citep[cf.,][]{Gillett:73,
Geballe:85, Cohen:co:86}.  These features have been detected and
studied in a large number of stellar sources -- planetary nebulae
(PNe), reflection nebulae (RNe), \HII regions, Young Stellar Objects
(YSO) and galaxies -- indicating that the emitters of these features
are surprisingly widespread and extremely stable, and therefore form a
very important component of the interstellar medium.

The UIR features are generally attributed to Polycyclic Aromatic
Hydrocarbons (PAHs) or PAH related molecules
\citep{Allamandola:rev:89, Puget:revpah:89}, although the exact
molecular identification remains uncertain. Compelling evidence
suggests that the features are due to a complex mixture of ionized and
neutral, possibly substituted/complexed PAHs of many different sizes
(\cite{Schutte:model:93, Bakes:photoelec:94, Boulanger:lorentz:98,
Bakes:modelI:01, Hony:oops:01}, hereafter HVP01,
\cite{Verstraete:prof:01, Pech:prof:01, Peeters:prof6:02}, hereafter
PHV02, \cite{Hudgins:colorado:04}).  Therefore the overall appearance
of the features including the profiles, relative strengths and peak
positions, are determined by a large number of PAH related parameters,
which are determined by local conditions and the PAH history.

High-resolution spectroscopy, as obtained with the \textit{Infrared
Space Observatory} \citep[ISO,][]{Kessler:iso:96} and with current
ground-based IR instruments, has revealed a richness of widespread
variations in the relative strength and profiles of the features from
source to source and within sources (\citet{Verstraete:m17:96,
Tielens:parijs:99, Maillard:99,Joblin:isobeyondthepeaks:00}; HVP01;
\citet{Vermeij:pahs:01}; PVH03: \citet{Goto:03, Madden:colorado:03,
Song:33:03, Song:colorado:03, Bregman:04, Peeters:review:04,
Peeters:pahtracer:04}; Galliano et al., in prep.).  PHV02 have
classified the sources in their sample according to the peak positions
of the CC modes at 6.2, 7.7 and 8.6 $\mu$m independently (Fig.~17 of
PHV02 and Fig.~\ref{chvscc_fig:var}, panels b and c, this paper).
Interestingly, correlations in the features' profiles were found and
three main classes with significantly different feature profiles were
identified. Most sources in their sample are in class $\mathcal{A}$
($A_{6-9}$ hereafter).  Sources with features peaking at significantly
longer wavelengths were referred to as class $\mathcal{B}$ ($B_{6-9}$
hereafter). The two sources in their sample showing a 6.3 $\mu$m
feature exhibit neither a 7.7 $\mu$m complex nor a 8.6 $\mu$m feature.
Instead, both sources show a broad emission feature at 8.22 $\mu$m.
These sources are referred to as class $\mathcal{C}$ ($C_{6-9}$
hereafter). Furthermore, the observed 6--9 $\mu$m PAH spectrum was
found to depend on the type of object and linked to local physical
conditions.

To extend our insight of the PAH family and its relationship with the
local physical conditions, we investigated the profiles of the CH
modes at 3.3 and 11.2 $\mu$m in the ISO-SWS spectra of a wide variety
of sources.  In Sect.~\ref{chvscc_DATA} our sample and the
observations are presented; the data reduction, the decomposition of
the spectra and the influence of extinction are discussed. In
Sect.~\ref{chvscc_PROF}, the 3.3 and 11.2 $\mu$m features are first
classified according to their profile and the relations between the
classifications of all major features in the 3--12 $\mu$m range are
investigated, as well as their correlation with object type.
Subsequently, variations in features associated with CH and CC modes
are compared. In Sect.~\ref{chvscc_spectroscopy}, the spectral
characteristics of PAHs in the 3--12 \mum\, range as measured in the
laboratory and calculated by quantum chemical theories are summarized.
The implications of the spectroscopic and observational conclusions
made in this paper are discussed in Sect.~\ref{chvscc_discussion}.
Finally, in Sect.~\ref{chvscc_Concl} our main results are summarized.
 
\section{The data}\label{chvscc_DATA}

\subsection{The sample}\label{chvscc_Sample}
For this study, we take the sample of PHV02, which included 57
different objects, all showing UIR bands, and reduce it to those
spectra with sufficient signal-to-noise (S/N) in the 3.3 and/or 11.2
$\mu$m regions (Table \ref{chvscc_tab:sample}).  In addition, we
excluded HD~100546 which exhibits crystalline silicates
\citep{Malfait:crystsil} and hence the 11.2 \mum\, emission band is a
combination of PAH emission and crystalline silicate emission. This
reduced sample includes 49 sources from a wide variety of objects,
ranging from RNe, (compact) \HII regions, YSO, Post-AGB stars, PNe to
galaxies.

\subsection{Observations and reduction}\label{chvscc_OBS}

All spectra in the sample were obtained with the Short Wavelength
Spectrometer \citep[SWS,][]{deGraauw:sws:96} on board ISO, using the AOT01
scanning mode at various speeds or the AOT06 mode, with a resolving
power ($\lambda / \Delta\lambda$) of $\sim$500-1500.

The data were processed with the SWS Interactive Analysis package
IA$^3$ \citep{deGraauw:sws:96} using calibration files and procedures
equivalent with pipeline version 7.0 or later. Further data processing
consisted of bad data removal and rebinning with a constant
resolution, as described in \citet{Peeters:cataloog:01}.

In case of high fluxes, the spectra can suffer from memory effects.
Here, this only applies to the 11.2 $\mu$m feature. At the time the
data reduction was done, no memory correction tool was available and
the average of the up and down scans is taken. The influences of
memory effects is investigated as described in
PHV02 and are
found not to alter the 11.2 $\mu$m feature profile significantly.

\clearpage

%%%%%%%%%%%%%%%%%%%%%%%%%%%%%%%%%%%%%%%%%%%%%%%%%%%%%

\begin{table*}[!htb]
\caption{The classifications of the 3--12 $\mu$m PAH features associated with each object.}
\label{chvscc_tab:sample}
\scriptsize
\begin{center}
\begin{tabular}{lc@{\hspace{6pt}}c@{\hspace{6pt}}c@{\hspace{6pt}}c@{\hspace{6pt}}cl}
 & & & & \\[-.2cm]
\hline  \hline \\[-8pt]
\multicolumn{1}{c}{Source}    & \multicolumn{5}{c}{PAH features} & Object type\\
 & 3.3$^a$ & 6.2$^{a,b}$ & 7.7$^{a,b}$ & 8.6$^{a,b}$ &11.2$^a$ & \\[2pt]
  \hline \\[-7.pt]
IRAS 10589-6034$^{\dagger?}$       & A   & A & A & A    & A &  compact H~{\sc ii} region\\ 
IRAS 12063-6259$^{\dagger}$        & A   & A & A & A    & A &  compact H~{\sc ii} region \\
IRAS 15384-5348$^{\dagger}$        & A   & A  & A & A   & A & compact H~{\sc ii} region\\ 
IRAS 15502-5302$^{\dagger}$        & A   & A  & A & y   & y &  compact H~{\sc ii}  region\\
IRAS 18116-1646$^{\dagger?}$       & A   & A  & A  & A  & A & compact H~{\sc ii} region\\
IRAS 18032-2032$^{\dagger}$        & A   & A  & A & A   & A & compact H~{\sc ii} region\\ 
IRAS 18317-0757$^{\dagger}$        & A   & A  & A & A   & A & compact H~{\sc ii} region\\
IRAS 18434-0242$^{\triangle,\dagger}$ & A   & A  & A & $\natural$ & A&
compact H~{\sc ii} region\\
IRAS 19442+2427$^{\triangle,\dagger}$ & A  & A  & A & A & A & compact H~{\sc ii} region\\  
IRAS 21190+5140                     & A  & A  & A & A   & A & compact H~{\sc ii} region\\  
IRAS 22308+5812                     & A  & A  & A & A   & A & compact H~{\sc ii} region\\
IRAS 23030+5958                     & A  & A  & A & A   & A & compact H~{\sc ii} region\\ 
IRAS 23133+6050                     & A  & A  & A & A   & A & compact H~{\sc ii} region\\  
W3A 02219+6125$^{\triangle,\dagger}$ & A  & A  & A & A   & A & compact H~{\sc ii} region\\
OrionBar D2   & B1 & A  & A  & A & A   &   H~{\sc ii} region\\
OrionBar D5   & A & A  & A  & A & A   &   H~{\sc ii} region\\
OrionBar D8   & A & A  & A  & A & A   &   H~{\sc ii} region\\
OrionBar H2S1 & A & A  & A  & A & A   &   H~{\sc ii} region\\
OrionBar BRGA & A & A  & A  & A & A   &   H~{\sc ii} region\\
Orion PK1$^{\dagger,\flat}$               & A & A  & A  & A & A   &  H~{\sc ii} region\\
Orion PK2$^{\dagger,\flat}$               & A & A  & A  & A & A   &  H~{\sc ii} region\\
G327.3-0.5$^{\dagger}$              & A & A/B1 & A & A & A   & H~{\sc ii}  region\\
IRAS 02575+6017$^{\dagger,\flat}$& A    & A & A & A    & A &   H~{\sc ii} region +YSO\\  
GGD-27 ILL$^{\dagger,\flat}$  & A$^V$   & A  & A & A   & A & star forming region\\
S106 (IRS4)$^{\dagger,\flat?}$& A  & A  & A & A & A    & YSO\\
IRAS 03260+3111               & A  & A  & A & A   & A    & Herbig AeBe star\\
BD+40 4124                    & A  & A  & A & A   & A     & Herbig AeBe star\\
CD-42 11721$^{\dagger?}$      & A  & A  & A & A   & A     & Herbig AeBe star\\
CD-42 11721(off)$^{\dagger?}$ &A  & A   & A & A   & A     &   Herbig AeBe star\\
HD 97048$^{\triangle}$        & A  & A   & AB & A   & A     & Herbig AeBe star\\
HD 179218                     & B$^V$ & B1/2& B & B& y& isolated Herbig AeBe star\\
NGC 7023 I                    & A & A    & A & A & A    & RN  \\
NGC 2023                      & A & A    & A & A & A    & RN  \\
MWC 922$^{\triangle}$         & B1 & B1  & A & $\natural$ & y & emission-line star\\
CRL 2688                      & A  & C   & C & C & y    & Post-AGB star\\
HD 44179$^{\triangle}$        & B2 & B2  & B & B & B    & Post-AGB star\\
HR 4049$^{\diamond}$          & B  & B2  & B & B & B    &  Post-AGB star\\
IRAS 16279-4757               & A  & A   & A & A & A    & Post-AGB star\\  
IRAS 13416-6243$^{\diamond}$  & B  & C   & C & C & y    & Post-AGB star\\[5pt]
IRAS 21282+5050               & A  & A   & AB& A & A(B) & PN\\
BD+30 3639$^{\triangle}$      & A  & B1  & B & B & A(B) &  PN\\
Hb5                           & A  & A   & $\triangleleft$  & A & B    & PN \\
He 2-113                      & A  & B2  & B & B & B    & PN \\
IRAS 07027-7943               & A  & B2/3& B & B & y    & PN \\
IRAS 17047-5650$^{\triangle}$ & A  & B1  & B & B & B  & PN \\
NGC 7027$^{\triangle}$        & A  & A   & B & B & A(B) &  PN \\   
circinus$^{\dagger}$          & y  & A   & A & A & B    & Seyfert 2 galaxy \\
M 82$^{\dagger?,\flat}$       & A  & A   & A & A & A(B) &  Starburst galaxy \\
NGC 253$^{\flat}$             & A  & A   & A & A & A(B) & Seyfert galaxy\\[5pt]
  \hline \\[-.55cm]
\end{tabular}
\end{center}
\scriptsize{
$^a$: Central wavelength is given in \mum; 
$^b$: from PHV02.$\;$
y: Feature is present but too weak and/or too noisy to classify; $\;$
$^\natural$: Source with an unusual 8.6 \mum\, feature
\citep[see PHV02,][]{Peeters:sc18434:04};   $\;$ 
$^\triangle$: Sources suffering from memory effects at
11 \mum;  $\;$   
$^\dagger$: Silicate absorption (9.7 \mum) present; $\;$ 
$^{\flat}$: Water ice absorption around 3 $\mu$m\, present; $\;$ 
$^\diamond$: Up- and down-scans disagree around 3.3 $\mu$m feature, but profile clearly of class $B$; $\;$
$^\triangleleft$: Strong [NeVI] present on top of the 7.6 \mum \,
feature hampers the classification; $\;$
$^V$: Classification by \citet{VanKerckhoven:phd:02}.}
\end{table*}
%%%%%%%%%%%%%%%%%%%%%%%%%%%%%%%%%%%%%%%%%%%%%%%%%%%

\clearpage

\subsection{The continuum}\label{chvscc_Cont}

To study the feature profiles in different sources, we subtract a
local continuum. This continuum determination is somewhat arbitrary. A
local spline continuum or a polynomial of order 2 is fitted to the
spectra in a wide wavelength region around the features
(Fig.~\ref{chvscc_fig:cont}). In the 3.3 $\mu$m region, the resulting
plateau underlying the 3.3 \mum\, PAH band, from about 3.2 to 3.6
$\mu$m, was subtracted after the subtraction of the continuum, using a
Gaussian with peak position 3.42 $\mu$m and a Full Width at Half
Maximum (FWHM) of 0.242 $\mu$m. For the 11.2 $\mu$m feature, the
plateau and the continuum are fitted together using a single spline.
For a review on the plateaus which underly many of the UIR bands, see
e.g. \citet{Peeters:review:04}. To assess the sensitivity of the resulting
profiles to the continuum or plateau choice, two extreme fits were
defined and subtracted. In general, the influence of the continuum
and/or plateau determination on the profile is very small and hence
does not change the classification/profiles of the sources.

\subsection{Extinction}\label{chvscc_Ext}

The data are not corrected for extinction. The influence of extinction
on the 3.3 and 11.2 $\mu$m feature profiles is investigated as
described in PHV02. The extinction is found to be
quite gray over the short wavelength regions and thus has no
significant influence on the profiles of the features.

\begin{figure}[t!]
    \centering
\epsscale{0.5}
\plotone{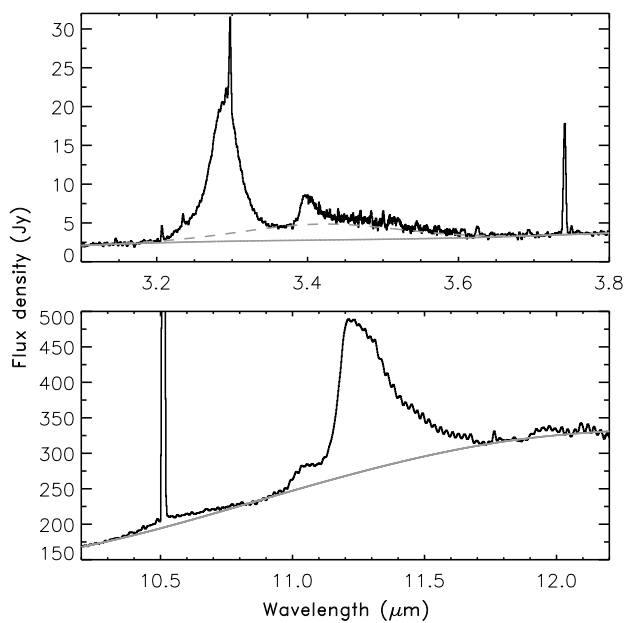}
    \caption{Illustrative examples of the continuum underneath the 3.3
      and 11.2 \mum\, features exemplified by NGC~7027. The full line
      represents the local continuum and the dashed line in the top
      panel represents the continuum after removing the plateau. The
      low-amplitude, high-frequency modulations between 11.2 and
      12.2~$\mu$m are an instrumental artifact due to
      interference. See Sect.~\ref{chvscc_Cont} for details.}
    \label{chvscc_fig:cont}
\end{figure}

Seven sources show water ice absorption around 3 $\mu$m (see Table
\ref{chvscc_tab:sample}). The 3.3 $\mu$m feature is situated in the
red wing of the water band and, hence, its profile could be
influenced. By drawing a local spline continuum
(Fig.~\ref{chvscc_fig:cont}), we compensate to some extent the effect
of the ice absorption and in the following we assume the ice
absorption has no influence on the profile that we derive. This is
corroborated by the fact that in our analysis those sources that
exhibit clear ice absorption do not show any systematic deviations in
the 3.3 \mum\, emission band.

\clearpage

\begin{figure}[tbh!]\centering
    \centering

\plotone{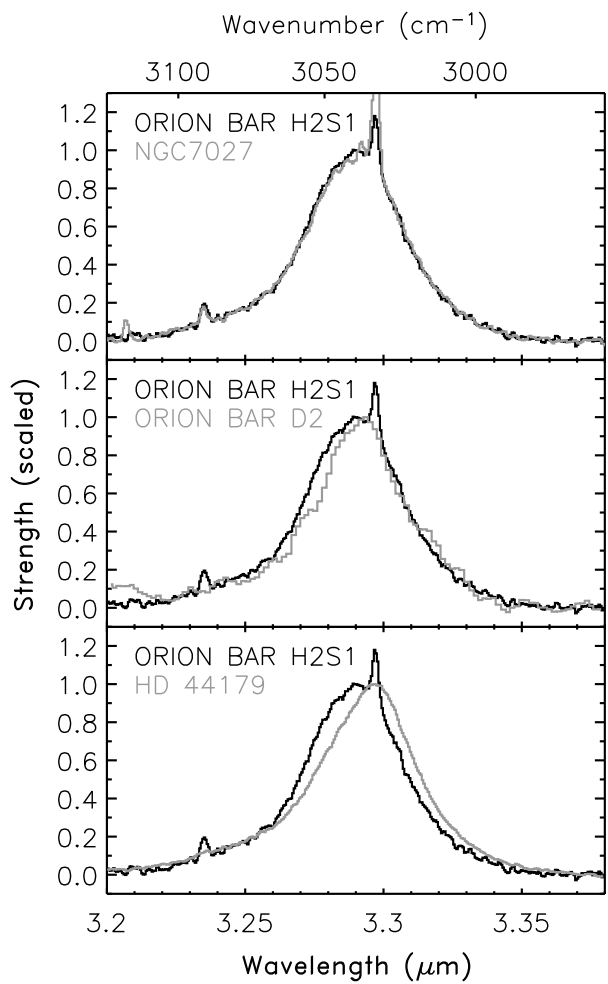}
    \caption{\small{Some examples of the 3.3 $\mu$m feature. The
features are normalised to the peak intensity. Note that a H
recombination line (Pf$\delta$ at 3.296 $\mu$m) is present in the
spectrum of the Orion Bar H2S1 and NGC~7027. The top-panel shows the shape
of the class $A_{3.3}$ features for two quite different
sources. This profile is observed in most sources. The middle- and
bottom-panel show the difference between class $A_{3.3}$ and,
respectively, $B1_{3.3}$ (represented by Orion Bar D2) and $B2_{3.3}$
(represented by HD~44179).}}
    \label{chvscc_fig:var33}
\end{figure}

\clearpage

\section{The feature profiles}\label{chvscc_PROF}

\subsection{The 3.3 $\mu$m feature}\label{chvscc_33Prof}

Nearly all sources in our sample show a pronounced 3.3 $\mu$m feature.
The similarity of most observed 3.3 $\mu$m feature profiles is very
striking. The top panel in Fig.~\ref{chvscc_fig:var33} shows this
profile for two quite different object types, the Orion Bar H2S1 (HII
regions) and NGC7027 (PN). Only six sources in our sample show
deviations in their 3.3 $\mu$m features (see
Table~\ref{chvscc_tab:sample} and Fig.~\ref{chvscc_fig:var33}, middle
and bottom panels).

Based on the profile of the 3.3 $\mu$m feature, the sources can be
classified into three classes, designated as $A_{3.3}$, $B1_{3.3}$ and
$B2_{3.3}$ (Tables~\ref{chvscc_tab:sample} and
\ref{chvscc_tab:classProp}). Class $A_{3.3}$, containing most sources,
has a symmetric profile with a peak position of $\sim$3.290 $\mu$m and
a FWHM of 0.040 $\mu$m (Fig.~\ref{chvscc_fig:var33}, top panel). Class
$B1_{3.3}$ and $B2_{3.3}$ have asymmetric profiles with peak positions
of $\sim$3.293 and $\sim$3.297 $\mu$m respectively, both with a FWHM
of 0.037 $\mu$m. The profile of class $B2_{3.3}$, represented by
HD~44179, is redshifted with respect to class $A_{3.3}$, and has a
less steep blue wing (Fig.~\ref{chvscc_fig:var33}, bottom panel). The
profile of class $B1_{3.3}$, represented by the Orion Bar D2, is intermediate;
its red wing is similar to the $A_{3.3}$ profile, but the blue wing is
clearly shifted to longer wavelengths and the peak position is
slightly to the red of the $A_{3.3}$ profile
(Fig.~\ref{chvscc_fig:var33}, middle panel).  The up- and down-scans
of HR~4049 and IRAS~13416-6243 disagree around the peak and the
spectrum of HD~179218 is noisy. Therefore, we can not distinguish
between class $B1_{3.3}$ or $B2_{3.3}$. However, we can say their
profiles do not resemble the class $A_{3.3}$ profile but belong to
either $B1_{3.3}$ or $B2_{3.3}$, indicated as $B_{3.3}$.

\subsection{The 11.2 $\mu$m feature}\label{chvscc_112Prof}

As a consequence of the selection requirements in Sect.\ref{chvscc_Sample},
the 11.2 \mum\, band profile is not studied for the class $C_{6-9}$
sources and only 5 sources of class $B_{6-9}$ are included in this
sample (Table~\ref{chvscc_tab:sample}).

\begin{figure}[!t]
    \centering

\plotone{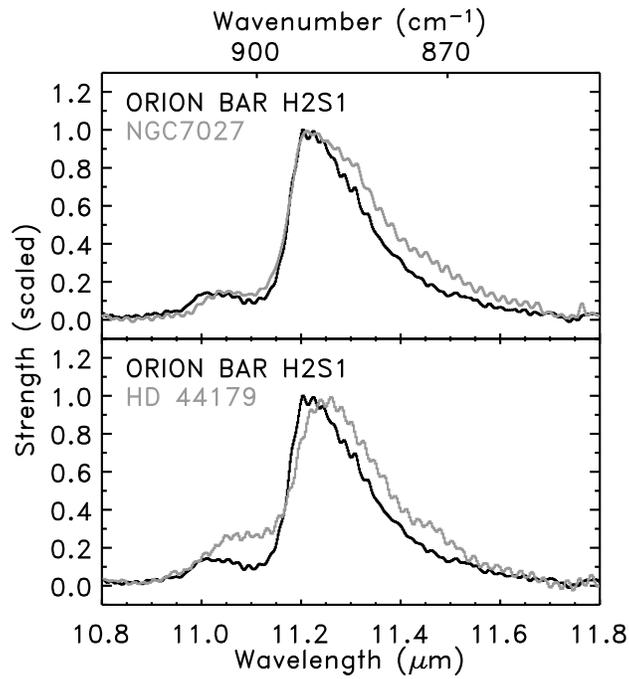} 
   \caption{Some examples of the 11.2 $\mu$m feature. The features
are normalised to the peak intensity. The top-panel shows the class
$A_{11.2}$ feature, represented by the Orion Bar H2S1 and the class
$A(B)_{11.2}$ feature, represented by NGC~7027. The difference between
class $A_{11.2}$ and $B_{11.2}$ (represented by HD~44179) is
illustrated in the bottom panel.}
    \label{chvscc_fig:var112}
\end{figure}

All sources in this reduced sample show a pronounced 11.2 \mum\,
feature, sometimes preceded by a weak feature at about 11.0 \mum. The
emission profiles of the 11.2 \mum\, feature are distinctly asymmetric
with a steep blue rise and a red tail.  A definite range in peak
positions and in the strength of the red wing when normalized so that
the profile's peak intensity equals 1 (denoted as tail-to-top ratio)
is present in our sample. Perusing the derived profiles, we recognize
2 main classes, which we will designate by $A_{11.2}$ and $B_{11.2}$
(Table~\ref{chvscc_tab:classProp}, Fig.~\ref{chvscc_fig:var112}).
First, the majority of the 11.2 \mum\, bands peak between 11.20 and
11.24 \mum, show a very steep blue rise, and have a low tail-to-top
ratio (Fig.~\ref{chvscc_fig:var112}, top panel, black profile). This
group will be referred to as class $A_{11.2}$. Within this class, the
peak position and the top of the profiles vary slightly.  Class
$B_{11.2}$ represents sources with a peak position of $\sim$11.25
\mum, a less steep blue rise (compared to class $A_{11.2}$) and a high
tail-to-top ratio (Fig.~\ref{chvscc_fig:var112}, bottom panel).  A few
sources (class $A(B)_{11.2}$) clearly belong to class $A_{11.2}$ in
term of peak position and steepness of the blue wing, but have a
tail-to-top ratio similar to class $B_{11.2}$
(Fig.~\ref{chvscc_fig:var112}, top panel, gray profile).

\clearpage

%%%%%%%%%%%%%%%%%%%%%%%%%%%%%%%%%%%%%%%%%%%%%%%%%%%%%%%%%%%%%%%%%%%%%%

\clearpage

%%%%%%%%%%%%%%%%%%%%%%%%%%%%%%%%%%%%%%%%%%%%%%%%%%%%%%%%%%%%%%%%%%%%%%
\begin{table*}[!bth]\centering
\caption{Overview of the defined classifications of the 3.3 and 11.2
  $\mu$m PAH features. See text for details.}
   \begin{center} 
\begin{tabular}{l c c c c c l}
  \hline \\[-5.pt]
  Class & & \multicolumn{2}{c}{peak wavelength} & & FWHM  &
  \multicolumn{1}{c}{profile} \\[2pt] 
  \cline{3-4}\\[-7pt]
        & & $\mu$m      & cm$^{-1}$             & & $\mu$m &        \\ [5pt] 
  \hline \\[-5.pt] 
  $A_{3.3}$  & & $\sim$3.290 & $\sim$3039 & & $\sim$0.040 & symmetric \\ 
  $B1_{3.3}$ & & $\sim$3.293 & $\sim$3037 & & $\sim$0.037 & asymmetric\\ 
  $B2_{3.3}$ & & $\sim$3.297 & $\sim$3033 & & $\sim$0.037 & asymmetric\\[-5pt] 
  \multicolumn{7}{c}{\dotfill} \\ 
  $A_{11.2}$ & & $\sim$11.20-11.24 & $\sim$889.6-892.9 & & $\sim$ 0.17 & asymmetric\\ 
  $A(B)_{11.2}$& &$\sim$11.20-11.24& $\sim$889.6-892.9 & & $\sim$ 0.21 & asymmetric\\
  $B_{11.2}$ & & $\sim$ 11.25 & $\sim$888.9 & & $\sim$ 0.20 & asymmetric\\[5pt]
  \hline \\[-5pt] \\ \\
\end{tabular}
\label{chvscc_tab:classProp}
  \end{center} 
\end{table*}
%%%%%%%%%%%%%%%%%%%%%%%%%%%%%%%%%%%%%%%%%%%%%%%%%%%%%%%%%%%%%%%%%%%%%%

\clearpage

\begin{figure*}[!bht]
    \centering
\epsscale{0.9}
\plotone{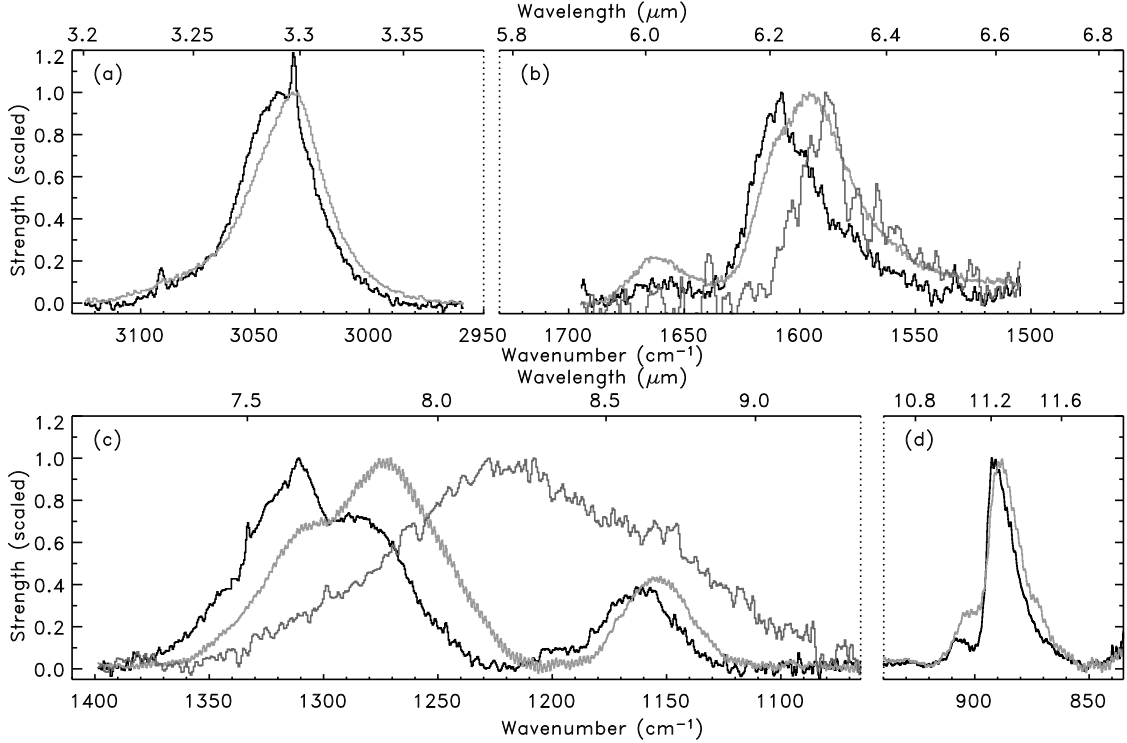}
\caption{An overview of the observed variations in the profiles of the
  3--12 $\mu$m PAH features. The spectra in each panel are normalised
  independently to the peak intensity. For all features, class $A$
  peaks at the shortest wavelengths (black line), and class $B$ peaks
  at even longer wavelength. In the 6--9 $\mu$m region, another class
  ($C$) is defined peaking at even longer wavelength and having a
  deviating spectral appearance (see Sect.~\ref{chvscc_INTRO} and
  PHV02 for details). Class $A$ is represented by IRAS~23133+6050 in
  the 6--9 \mum\, region and by the Orion Bar H2S1 for the 3.3 and
  11.2 \mum\, features. HD~44179 represents class $B$ (classes $B2_{3.3}$ and
  $B2$ for the 3.3 and 6.2 PAH bands respectively and class $B$ for the 7.7,
  8.6 and 11.2 \mum\, PAH bands) and IRAS~13416-6243 represents class
  $C$. Note that H recombination lines (Pf$\delta$ at 3.296 $\mu$m and
  HI(6-5) at 7.458\mum\,) are present in the spectrum of the Orion Bar
  H2S1 and IRAS~23133+6050. }
\label{chvscc_fig:var}
\end{figure*}
%%%%%%%%%%%%%%%%%%%%%%%%%%%%%%%%%%%%%%%%%%%%%%%%%%%%%

\clearpage

\subsection{Feature profile correlations}\label{chvscc_Corr}

Comparing the classifications of the main PAH bands (Table
\ref{chvscc_tab:sample}), we find i) 34 (out of 49) sources where all
classified features belong to either class A or class B or class C, of
which 31 are class A and 3 are class B, ii) 6 sources where all
classified features except the 3.3 \mum\, feature belong to either
class A or class B or class C and iii) 2 sources where all classified
features except the 11.2 \mum\, feature belong to either class A or
class B or class C. Concluding, we can see that in this large sample
of sources, the longer wavelength feature profiles correlate slightly
better with each other than with the 3.3 \mum\, feature.

To illustrate the observed variations in the class $B$ profile
correlations, a few examples can be given. HD~44179 and HR~4049 have
class $B$ profiles, considering all features. BD+30~3639, however, has
6--9 $\mu$m features of class $B$ and 3.3 and 11.2 $\mu$m features of
class $A$.  Other sources, as IRAS~17047-5650 and He~2-113, have 3.3
$\mu$m features of class $A_{3.3}$, but class $B$ for the longer
wavelength features. Of the two sources of class $C_{6-9}$, one
(IRAS~13416-6243) has a 3.3 $\mu$m feature of class $A_{3.3}$ and the
other (CRL~2688) of class $B_{3.3}$. In contrast, within the 6--9
$\mu$m range, spectral class $B_{7.7}$ also (almost) invariably
implies $B_{6.2}$ and $B_{8.6}$ and vice versa.  The 11.2 $\mu$m
classes follow, in this respect, more the 6--9 $\mu$m trend than
the 3.3 $\mu$m behaviour.\\

Perusing the spectra, it is striking that the variations in the 6--9
$\mu$m region are much more pronounced than those of the 3.3 or 11.2
$\mu$m features (cf., Fig~\ref{chvscc_fig:var}; Table
\ref{chvscc_tab:classProp} and Table 3 of PHV02).  Specifically the
maximum peak position separation in wavenumber space varies in going
from class A to B by about 20, 80 and 15 cm$^{-1}$ for the 6.2, 7.7
and 8.6 $\mu$m features, respectively, and only by about 6.5 and 4.0
cm$^{-1}$ for the features at 3.3 and 11.2 $\mu$m, respectively.\\

From an astronomical point of view, correlations between the classes
and the different object types are also found. Almost all \HII
regions, non-isolated Herbig AeBe stars and YSO's in our sample are in
class $A$, considering all features. An interesting exception is the
Orion Bar position D2, showing a $B1_{3.3}$ feature and class $A$
6--12 $\mu$m features. The galaxies in our sample are in class $A$,
considering all features, except for the 11.2 $\mu$m feature, which is
of class $A(B)_{11.2}$ or $B_{11.2}$.  Most PNe have 6--9 $\mu$m
features of class $B_{6-9}$, a 3.3 $\mu$m feature of class $A_{3.3}$
and a 11.2 $\mu$m feature of class $B_{11.2}$ or $A(B)_{11.2}$. The PN
Hb5 has a 11.2 $\mu$m feature of class $B_{11.2}$ while the other
bands belong to class $A$.  The Post-AGB stars in our sample are
spread over the classes. IRAS~16279-4757 is in class $A$ considering
all features.  IRAS~21282+5050 has a 11.2 $\mu$m feature of class
$B_{11.2}$ while the other bands belong to class $A$.  The two
extreme metal poor binary system Post-AGB stars, HD~44179 and HR~4049,
are in class $B$ considering all features. The same is true for the
isolated Herbig AeBe star HD 179218, although its 11.2 $\mu$m feature
is too weak/noisy to be classified. Thus, in conclusion, the
correlation between profile class and object type is much tighter in
the 6--9 $\mu$m region than in the 3.3 and 11.2 $\mu$m
regions. \\

\section{The infrared emission features and PAHs}
\label{chvscc_spectroscopy}

The 3.3 and 11.2 $\mu$m PAH bands arise from the radiative relaxation of
CH stretching and CH out-of-plane bending modes of highly
vibrationally excited polycyclic aromatic hydrocarbons.  This is in
contrast with the nature of the 6.2 and 7.7 \mum\, PAH features which
originate from vibrations mainly involving CC stretching motions.
Thus, while one would expect strong global correlations amongst all
of the bands that arise from the same family of molecules, when
comparing the behavior of bands involving CH modes with those
involving CC modes, subtle but important differences in behavior are
expected.  These differences reflect not only differences in local
conditions, but also structural differences within the emitting
family itself, and in the case of the 3.3 $\mu$m band emission from a
distinctly different subset of the emitting PAH population.

While discussing the dependence of the precise position of any
vibrational mode, one should keep in mind that the vast majority of
the laboratory data available to address each of these bands has been
measured in absorption, not emission.  In the astrophysical case,
where PAH spectra are measured in emission instead of absorption, the
extent of molecular excitation also influences peak position and
profile.  The interstellar spectrum arises from the combined emission
of a complex mixture of vibrationally excited PAHs.  In emission, each
individual line should have an approximately 30 cm$^{-1}$ FWHM,
consistent with the natural linewidth expected from each emitting
molecule \citep{Allamandola:rev:89}.  This natural linewidth arises
from intramolecular vibrational energy redistribution, not the
blending of individual rotational lines in the emitting molecules
which remain rotationally cool throughout the excitation/emission
process \citep[e.g.][]{Brenner:benz+naph:92,Cook:excitedpahs:98}.  Due
to the high internal energy content of the emitting molecules, a 10 to
as much as 40 cm$^{-1}$ peak position redshift is intrinsic to the
emission process depending on the internal excitation and the emitting
feature wavelength \citep{Flickinger:91, Brenner:benz+naph:92,
Colangeli:T:92, Joblin:T:95, Cook:uvlaserexpahs:98, Wagner:2000}.

\subsection{The CH stretching mode and the 3.3 \mum\, band}
\label{chvscc_spectroscopy_ch}

As with all fundamental vibrational frequencies, the precise position
of the 3.3 \mum\, band depends on many factors including charge state,
molecular size, structure, molecular heterogeneity, and so on.  Each
of these factors which influence fundamental band position are
discussed below.

\begin{figure}[!t]
    \centering
\epsscale{0.5}
\plotone{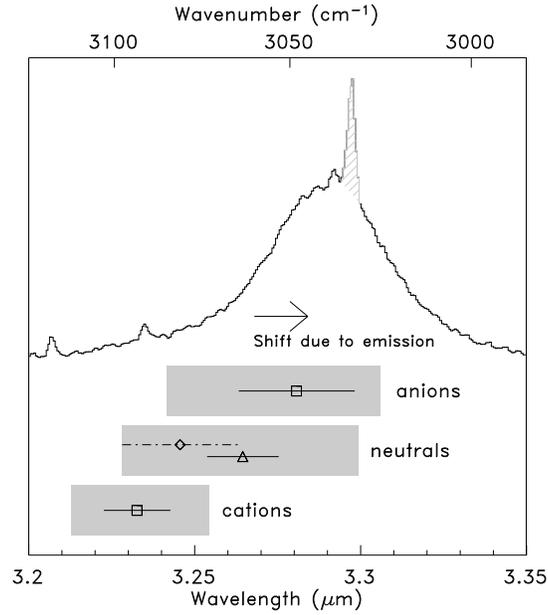}
    \caption{Comparison of the interstellar emission feature from
NGC~7027 (class A$_{3.3}$) to the charge specific CH stretch regions
for the entire Ames sample. The striped, grey feature is
Pf$\beta$. The shaded areas represent, for each charge state, the full
measured range in positions for the CH stretch mode. The symbols
crossed by horizontal lines represent the average position and its
standard deviation. The diamond and triangle crossed by horizontal
lines are specific for large (N$_{C}$ $\ge$ 40) and small (N$_{C}$
$<$ 40) neutrals respectively. The arrow represent a 15 cm$^{-1}$
peak position redshift intrinsic to the emission process (see text for
details).  }
    \label{lab33}
\end{figure}

\paragraph{Charge state :} 
The strongest constraint concerns charge state of the PAHs responsible
for the interstellar 3.3 $\mu$m emission band.  In the Astrochemistry
Laboratory at NASA Ames, we now have the experimental infrared spectra
of over 100 isolated PAHs in their neutral as well as positively
charged states.  This spectral database is comprised of PAHs with a
number of carbon atoms, N$_C$, between 10 and 50.  These data are
supplemented with theoretical spectra which extend the size to
N$_C$=96.  The number of PAHs for which anion spectra are available is
considerably smaller, currently 27.  Analysis of these spectra shows
that the CH stretching frequencies are sensitive to PAH charge as well
as size (Hudgins et al., in prep.).  This is illustrated in
Fig.~\ref{lab33}, which compares the interstellar emission feature
from NGC~7027 (class A$_{3.3}$) to the charge specific CH stretch
regions for the entire Ames sample.  Fig.~\ref{lab33} shows that the
CH stretch in PAH anions spans the range from 3085 to 3025 cm$^{-1}$
(3.24 to 3.30 $\mu$m), neutral PAHs from 3094 to 3042 cm$^{-1}$ (3.23
to 3.29 $\mu$m), and PAH cations from 3112 to 3073 cm$^{-1}$ (3.21 to
3.25 $\mu$m).  The symbol within each range indicates the average
position for the PAHs in the Ames database. These are 3048, 3081, 3063
and 3093 cm$^{-1}$ (3.28, 3.24, 3.26 and 3.23 $\mu$m) for the anionic
forms, large neutrals, small neutrals and the cationic forms
respectively.

PAH cations seem least likely to contribute to the 3.3 \mum\,
feature. First, Fig.~\ref{lab33} shows they require the largest
redshift possible (40cm$^{-1}$) to overlap only the blue wing of the
feature. Second, PAH cations have inherent weak emission in the 3.3
\mum\, region, i.e. the CH band of cations is suppressed by a factor 50
to 100 compared to that of neutral PAHs
\citep{Langhoff:neutionanion:96}. Lastly, the 3.3 and 11.2 bands are
well correlated and behave differently then the 6.2 and 7.7 bands
which are dominated by PAH cations \citep{Hony:oops:01}.  In contrast,
Fig.~\ref{lab33} shows that the CH stretches for neutral PAHs, when
redshifted due to the high internal energy content, fall directly
under the most intense portion of the interstellar 3.3 $\mu$m band
envelope while the PAH anions require a small redshift to fall
underneath the 3.3 \mum\, band envelope. The breadth and small
variance of the interstellar 3.3 $\mu$m emission band likely arises
because the CH stretching position (in absorption) for the vast
majority of isolated, neutral and anionic PAHs falls between 3085 to
3025 cm$^{-1}$.  With the 10 to 40 cm$^{-1}$ redshift inherent in
emission, this range becomes ~3075 to 2985 cm$^{-1}$ (3.25 to 3.35
$\mu$m), a range which straddles the wavelength region of this feature
listed in Table~\ref{chvscc_tab:classProp}.

\paragraph{Size :} 
PAH size is particularly important in the case of the interstellar 3.3
$\mu$m band since this feature originates from the smallest members of
the emitting PAH population \citep{Allamandola:rev:89,
Schutte:model:93}.  PAHs containing between roughly 25 to 70 carbon
atoms contribute most of the emission in this feature, with the
species between N$_C$ $\sim$ 25 to 50 dominant.  This is in contrast
to the situation for the longer wavelength emission features which
arise from increasingly larger, and overlapping members of the
interstellar PAH population.

Interestingly, within the size range of the current Ames sample, there
appears to be a bimodal distribution of the frequencies for the
neutral species (see Fig.~\ref{lab33}).  For the neutral PAHs, the CH
stretch for the smallest members of the sample is centered near 3060
cm$^{-1}$ (3.27 $\mu$m), while the larger members have frequencies
centered at 3090 cm$^{-1}$ (3.24 $\mu$m).
The origin of the bimodal distribution is not clear from the current
sample.

\clearpage

\clearpage

\paragraph{Edge Structure :} PAH edge structure is also important.  If the
periphery of an individual PAH is not regular, the local environment, that 
each CH bond samples, can perturb the individual stretching
frequencies.  Thus, for a large molecule with long straight edges
there is little interaction between the stretching motions of CH
groups on adjacent rings and the FWHM of the dominant peak in the
region is 18 cm$^{-1}$ (0.019 $\mu$m at 3.3 $\mu$m).  On the other hand, if a
similarly large molecule that has jagged edges with many ``notches" and
``bay regions", CH groups on opposing sides of these gaps interact,
perturbing their stretching motions.  As a result, the FWHM of the CH
stretching feature in such a species is substantially broadened and
may be as wide as 70 cm$^{-1}$ (0.074 $\mu$m) or more (Hudgins et
al., in prep.).

\paragraph{Heteroatom substitution :} \citet{Mattioda:npahs:03} have 
investigated the spectra of 11 N-substituted PAHs, spanning the N$_C$
range from 9 to 21, and found that nitrogen substitution does not influence
the CH stretching band frequency.  N substitution does influence other
PAH vibrational transitions \citep[PHV02;][Hudgins, Bauschlicher, \&
Allamandola, in prep.]{Mattioda:npahs:03}

\subsection{The CH out-of-plane bending modes}
\label{chvscc_spectroscopy_cc}

\begin{figure}[!t]
    \centering

\plotone{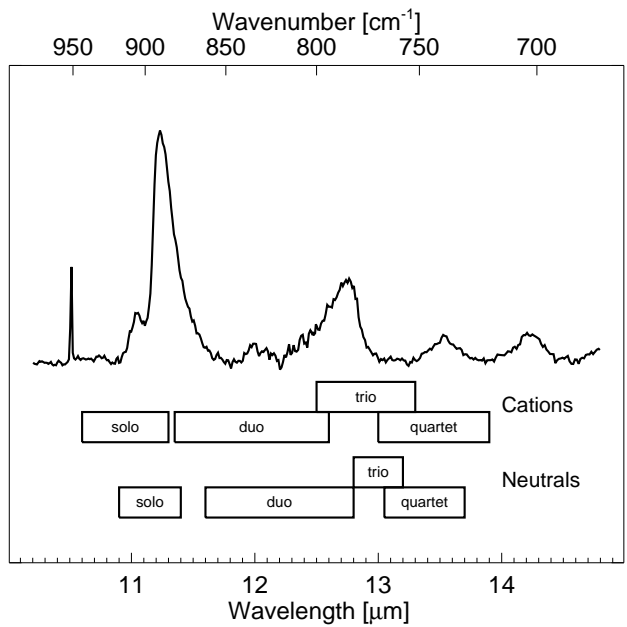}
    \caption{A comparison of the average interstellar spectrum (top)
      with the ranges characteristic of the CH out-of-plane bending
      modes of neutral and cationic PAHs (bottom). Details of the
      average interstellar spectrum are given in HVP01.  The boxes
      indicate the wavelength regions associated with the out-of-plane
      bending vibrations of the various adjacency classes of the
      peripheral CH groups as determined from matrix isolation
      spectroscopy. Aromatic rings carrying
      CH groups which have no neighbouring CH groups are termed
      ``non-adjacent'' or ``solo'' CH groups. Likewise, two adjacent
      CH groups are called ``duet'' CH's,
      three adjacent CH groups ``trio'' CH's and four adjacent CH groups
      ``quartet'' CH's. Figure taken from HVP01.  }
    \label{chvscc_oops}
\end{figure}

The aromatic CH out-of-plane bending features in the 11 to 15 \mum\,
spectral region are a good diagnostic for the classification of the
aromatic ring edge structures.  Indeed, the positions and intensities
of the bands in this spectral region reflect the type and number of
adjacent CH groups on the peripheral rings of the PAH structure
\citep[][HVP01]{Bellamy:58, Allamandola:autoexhaust:85, Cohen:85,
Leger:photothermodisoc:89, Roche:orion:89, Witteborn:89,
Allamandola:modelobs:99, Hudgins:tracesionezedpahs:99}.  Our current
understanding is illustrated in Fig.~\ref{chvscc_oops} which
schematically compares the average interstellar emission spectrum with
the wavelength regions associated with different CH adjacency classes
for neutral and ionised, isolated PAHs.  Inspection of this Figure
shows that, while the ranges for matrix-isolated neutral PAHs do not
differ substantially from those reported in the literature, ionisation
produces some notable changes in region boundaries
\citep{Hudgins:tracesionezedpahs:99}.  Taking into account the roughly
0.1 \mum\, red-shift in the peak wavelength of 11.2 \mum\, PAH band in
emission relative to their position in absorption, the following
conclusions are drawn \citep[][HVP01]{Hudgins:tracesionezedpahs:99} :
(i) The weak interstellar emission band sometimes observed on the
short wavelength side of the dominant 11.2 \mum\, band and peaking
near 11.0 \mum\, falls in the region attributable to the solo-CH modes
of PAH cations; (ii) The peak of the 11.2 \mum\, interstellar band
falls squarely within the region for the solo-CH modes of {\it
neutral} PAHs and at the extreme long wavelength end of the range for
solo cationic modes; (iii) The variable red wing of the interstellar
11.2 \mum\, band likely carries a contribution from the duet-CH modes
of PAH cations, particularly those with condensed structures whose
features tend to fall near the lower extreme of the characteristic
region shown in Fig.~\ref{chvscc_oops}.

\clearpage

\clearpage

\subsection{The CC stretching modes}

The CC stretching modes occur in the 6--9 $\mu$m range.  Within this
range, the band at the longer wavelengths have a mixed CC stretch and
CH in-plane-bending vibration character.  All CC modes are very weak
relative to the CH modes in neutrals, but become the dominant bands in
ions \citep{Szczepanski:nature:93, Langhoff:neutionanion:96,
Hudgins:cationsI:94, Kim:gasphasepyrenecation:01}.  Hence, we will
focus here on the CC modes in cations.  The peak position of the pure
CC stretch is influenced by molecular size, structure, and molecular
heterogeneity (see e.g. PHV02).  Extensive laboratory and theoretical
quantum mechanical studies have shown that a peak as blue as 6.2
$\mu$m does not occur for pure-C PAH cations in the (interstellar)
size range 20-100 C-atoms.  Substitution of N deep in the fused ring
system induces strong IR activity of the CC modes at the highest
frequencies within this range and bands around 6.2 $\mu$m do become
intense (PHV02, Hudgins, Bauschlicher \& Allamandola, in prep.).  This
is the result of the delocalisation of the charge across the entire
PAH as a result of incorporating N in the ring.  While there is
presently no corroborating experimental or theoretical evidence, a
similar behavior may occur for large PAHs with uneven and irregular
edge structures, PAH clusters and/or PAH complexes with metals.  For
the longer wavelength modes, both pure-C PAH cations as well as
N-substituted PAH cations can show prominent bands near the
interstellar 7.6 $\mu$m position.  The origin of the 7.8 $\mu$m band
is more enigmatic, however, and no good match with laboratory species
exists.

\section{Discussion}\label{chvscc_discussion}

From the wealth of medium resolution IR spectra, it is clear that the
major PAH features show clear variations in profile and peak
position. Large variations are found in profiles of the features in
the 6--9 \mum\, region while smaller ones are found in those of the
3.3 and 11.2 \mum\, features. Furthermore, the 3.3 \mum\, feature is
very invariant under highly variant physical conditions. The different
profile classes are linked with each other and with object type; this
correlation, however, is more apparent in the 6--9 \mum\, region than
in the 3.3 and 11.2 \mum\, regions.

\subsection{The observed trends}

%% aperture
\citet{Tokunaga:33prof:91} recognized two types of 3.3 $\mu$m features
based on spectra taken with a $2.7''$ aperture. The profile, peak
position and width of their type 1 agree well with class $A_{3.3}$
defined here. Their type 2 has a similar peak wavelength compared to
our classes $B_{3.3}$, but is much narrower, with a FWHM of
$\sim$0.020 $\mu$m. Most sources of their sample belong to Type 1
while Type 2 is associated with only 3 sources (HD~44179, Elias~1,
WL~16). The 3.3 \mum\, PAH band in the ISO-SWS spectra (obtained
through an aperture of $14''$x$20''$) of their Type 1 sources belongs
to class $A_{3.3}$ while those of HD~44179 and Elias~1 belongs to
class $B_{3.3}$ (see Table~\ref{chvscc_tab:sample} and
\citet{VanKerckhoven:phd:02} for Elias~1).  \citet{Kerr:rr:99} and
\citet{Song:33:03} showed, in a detailed spatial study of HD~44179,
that the 3.3 $\mu$m profile varies, going from Type 2 to Type 1 with
distance from the central star. Hence, the observed profile will
depend on the sampled area on the sky. In particular, the 3.3 $\mu$m
profile of HD~44179 in our sample is observed through an aperture
$14''$x$20''$ and thus has an averaged profile of both Type 1 and 2
profiles, thereby explaining its large FWHM compared to the profile
found by \citet{Tokunaga:33prof:91}, \citet{Kerr:rr:99} and
\citet{Song:33:03}. In addition, the Type 1 / class A$_{3.3}$ profile
is not unique to certain object types. Whether the Type 2 profile only
occurs in specific sources is currently unknown. However, few sources
are found that show a Type 2 or class B$_{3.3}$ profile. All except
one are post-AGB stars (3), isolated HAeBe stars (3) and an emission
line star (\citet{Molster:crystsilsample:02} classifies MWC~922 as
disk-like based upon the spectral energy distribution and its mm
continuum flux). Apparently, a circumstellar environment where dust
formation is on-going or a protostellar/protoplanetary environment is
an essential but not sufficient condition to have Type 2 or class
B$_{3.3}$ profiles. Indeed, some post-AGB stars or HAeBe stars show a
Type 1 / class A$_{3.3}$ profile. The fact that the 3.3~$\mu$m feature
varies considerably less than the other main features and mostly
exhibits the A$_{3.3}$ / Type 1 profile indicates that the conditions
that give rise to the B$_{3.3}$ and/or Type 2 profiles are more easily
lost than those that give rise to the B (and C) types for the other
bands. \citet[][2004 in prep.]{Song:colorado:03} and
\citet{Miyata:rr:04} also reported a varying 8.6 and 11.2 \mum\,
profile in HD44179, with a peak position going to shorter wavelengths
with distance from the star and hence going from class B to class
A. Note that also the 3.3 \mum\, feature also shifts bluewards with
distance from the star indicating that these three PAH feature are
correlated in this source. Similar as for the 3.3 \mum\, feature, the
observed profile will depend on the sampled area on the sky. In
addition, class A profile is not unique to certain object types.

%% Teff
\citet{Tokunaga:33prof:91} noted that the central stars of the few
sources showing a Type 2 profile, all had relatively low effective
temperatures ($\sim 10^4$ K), suggesting a correlation between stellar
effective temperature and PAH feature profiles. However, both profile
types are found within HD~44179 \citep{Kerr:rr:99, Song:33:03}. 
At the same time, some objects in our sample have a relatively low
stellar temperature, e.g. IRAS 03260+3111 (Jaschek et al. 1964), but
still show an A$_{3.3}$ class profile. This demonstrates that the
effective temperature cannot be the sole cause of the variations in
the 3.3~$\mu$m feature profile. Van Kerckhoven (2002) and PHV02 have
reached similar conclusions for the longer wavelength features.

%% G_0
Since variations are seen within objects, it may be that the local
radiation field, $G_{0}$, influences the profiles of the
feature. However, no link between $G_{0}$ and the different classes of
the 3.3 and 11.2 \mum\, features is found. Indeed, for both features
individually, the range in $G_0$ of the sample sources classified as
class $A$ straddles that of class $B$ and class $A(B)_{11.2}$
sources.  Analogously, PHV02 found no relation between $G_{0}$ and
profiles of the 6--9 \mum\, PAH emission features.  Hence, we can
conclude that G$_0$ alone cannot be the factor determining the
observed variation in the main PAH emission features from source to source
(observed through a large aperture).

\subsection{The origins of the variations in the 3.3 and 11.2 \mum\, CH emission features}
 
In the framework of the PAH hypothesis, the profiles and thus also the
variations in the profiles reflect the present molecular ensemble and
the excitation conditions. In the following, we examine if we can
isolate any effect which is clearly reflected in the observed
variations of the profiles.

 \subsubsection{Anharmonicity}

The profiles are very characteristic for anharmonic broadening in
highly vibrationally excited molecules \citep{Barker:anharm:87,
Kim:gasphasepyrenecation:01, Verstraete:prof:01, Pech:prof:01}.  Due
to cross-coupling between different modes, the peak position of an IR
band will shift towards lower frequencies.  Integration over the
energy cascade, leads then to a red shaded profile.  The red wing can
be further enhanced due to emission from vibrationally excited levels
(e.g. vibrational hot bands, $v= 2\rightarrow 1$) which are generally
shifted towards the red due to the anharmonicity of the vibrational
potential of the mode.  These anharmonicities can be quite different
for different modes.  Anharmonicity parameters for a few small PAHs
have been estimated from laboratory studies on the temperature
dependence of the peak wavelength and width of the absorption bands
\citep{Joblin:T:95} and from the difference between low temperature
absorption measurements and infrared multi-photon dissociation studies
of trapped PAH ions \citep{Oomens:03}.

The effects of anharmonicity have been studied theoretically
\citep{Verstraete:prof:01, Pech:prof:01}. These studies show that the
general characteristics of the interstellar 3.3, 6.2 and 11.2~$\mu$m
bands are well understood in the framework of highly excited
PAHs. However, these models cannot explain the large shift observed in
the 6.2 $\mu$m band between classes $A$, $B$, and $C$.  Such a shift
can only reflect a chemical/structural modification of the emitting
PAHs, for example due to the incorporation of N (PHV02; Hudgins,
Bauschlicher, \& Allamandola, in prep.).

The calculated 3.3 \mum\, profiles of \citet{Pech:prof:01} show that
an increase in excitation -- either through a decreased size or an
increase in internal energy -- shifts the peak position towards the
red and simultaneously increases the FHWM (note that the calculated
3.3 \mum\, profile has a larger asymmetry factor than observed in
space).  Our class $B_{3.3}$ profiles and the Type 2 profiles peak at
longer wavelengths but have smaller FWHM compared to class $A_{3.3}$ /
Type 1 profiles. Thus, the variation in profiles between classes
$A_{3.3}$ / Type 1 and either $B_{3.3}$ or Type 2 can not be due to
anharmonicity.

Perusing the \citet{Pech:prof:01} 11.2 \mum\, model profiles, we
notice that an increase in excitation -- either through a decreased
size or an increase in internal energy -- does increase the strength
of the red wing but can not shift the peak position or change the blue
rise.  In their study of the 11.2~$\mu$m band, \citet{Pech:prof:01}
focused on IRAS 21282+5050, a class A(B)$_{11.2}$ source and indeed
the A(B)$_{11.2}$ profile can be interpreted as originating from the
same PAH population that gives rise to the A$_{11.2}$ profile excited
to higher internal energy. In contrast, the difference between
A$_{11.2}$ and B$_{11.2}$, with the band as a whole displaced, are not
readily explained within the anharmonicity model.

\subsubsection{Composition}

The 3.3 and 11.2 \mum\, CH bands are likely due to the presence of a
family of PAH molecules and hence, the variation in peak position and
profile of the CH modes likely represents a small but significant
difference in the population of emitting species.  The variation in
the peak position of the 3.3 and 11.2 $\mu$m band, measured in the
laboratory or quantum-chemically calculated, is much larger (0.02
$\mu$m; 20 cm$^{-1}$ and 0.3 $\mu$m; 20 cm$^{-1}$ respectively) than
observed in space ($\sim$ 0.007 \mum; 6 cm$^{-1}$ and $\sim$ 0.05
$\mu$m; 4 cm$^{-1}$ respectively).  Although the convolution of the
size distribution with the excitation conditions can lead to a
different dispersion for the peak positions (compared to absorption
bands measured in the lab), it is likely that only a {\it small subset}
of the PAHs studied by these means contribute to the observed emission
from space.

\subsubsection{Molecular edge structure}

\paragraph{The 3.3 \mum\, feature :}

If we assume that the variation in the FWHM of the 3.3 \mum\, feature
between class A$_{3.3}$ / Type 1 and either B$_{3.3}$ or Type 2 is
mainly influenced by the molecular edge structure, this observed
variation in FWHM (and hence in class) implies that the PAH family
present in class B$_{3.3}$ and Type 2 sources are dominated by more
compact PAH molecules compared to the PAH family present in class
A$_{3.3}$ / Type 1 sources (Sect.~\ref{chvscc_spectroscopy_ch} and
Table~\ref{chvscc_tab:classProp}). Since the molecular structure of
the PAH molecules can be deduced based on the relative strengths of
the CH out-of-plane bending modes (HVP01) and assuming that the PAHs
responsible for the 3.3 \mum\, band have similar molecular structure
as those for the 11.2 \mum\, band (although these features arise from
the emission of a different subset of PAHs), we can test the
dependence of the FWHM of the 3.3 \mum\, feature on molecular
structure. Let's consider the following sources : HD~44179 belonging
to $B_{3.3}$ and $B_{11.2}$; NGC~7027, member of $A_{3.3}$ and
$B_{11.2}$ and the IRAS~18317 representing classes $A_{3.3}$ and
$A_{11.2}$.  HVP01 found that NGC~7027 is dominated by large, compact
PAHs with straight edges and IRAS~18317 by smaller or irregular
PAHs. Although these two sources represent the extremes in the
observed intensity ratio in their sample and hence the extremes in
molecular structure, they both belong to $A_{3.3}$. In addition, the
relative strengths of the CH out-of-plane bending modes of HD~44179
are comparable to NGC~7027 and hence their molecular structure is
similar (HVP01). However, their 3.3 \mum\, features belong to
different classes. Therefore, we can conclude that the molecular edge
structure is likely not the main cause of the variation in the FWHM of
the 3.3 \mum\, feature.

\paragraph{The 11.2 \mum\, feature :}

The little variance of the CH out-of-plane bending modes may be an
inherent property of interstellar PAHs.  Within the subset of PAHs
studied in the laboratory or quantum chemically, the peak position of
the solo modes in neutral and cation PAHs falls very closely to the
observed interstellar peak position of 11.2 $\mu$m, with the neutral
PAHs dominating the emission (see Fig. 8 in HVP01).  Now, possibly,
any molecular structure variations, which lead to shifts in the 6--9
$\mu$m CC bands, may be inherently accompanied by modification of the
CH peripheral structure which turn CH solo's into duo's or trio's.

Duet CH's of compact PAH cations may contribute to the red wing of the
11.2 \mum\, band
\citep[Sect.~\ref{chvscc_spectroscopy_ch}][HVP01]{Hudgins:tracesionezedpahs:99}. Hence,
it is possible that the observed variation in the 11.2 \mum\, band
reflects variation in the contribution of these cationic duet
modes. This would imply that $B_{11.2}$ and $A(B)_{11.2}$ objects have
more compact PAH cations with duets. Analysis of the pattern of
emission features in the CH out-of-plane bending modes region (i.e.
10--15 \mum, HVP01) reveals that evolved stars are dominated by large
compact structures while \HII regions are dominated by irregular
structures and hence is consistent with this conclusion.\\

\subsubsection{Carbon isotope effects}

Based upon laboratory studies, shifts in the peak positions of the
interstellar IR emission features have been attirbuted to $^{13}$C isotope
substitution \citep{Wada:13c:03}. We note that, while this might be
important in some extreme circumstellar environments (such as the Red
Rectangle), this cannot be a general explanation for the observed
shifts. First, not all the features shift in the same way at all
times. Second, we see pronounced variations in material which is
interstellar in origin and where isotope subsitution should be
minimal. The observed variations within a single \HII region or
reflection nebula present, of course, a case in point \citep[Joblin et
al. in prep.][]{Bregman:04}.

\bigskip 
To summarise, the variation in peak position is likely due to the
presence of a mixture of PAH molecules whose composition changes.
Both anharmonicity and molecular structure can explain variation in
the FWHM of the 11.2 \mum\, feature.  In contrast, the variation in
the FWHM of the 3.3 \mum\, feature between class A$_{3.3}$ / Type 1
and Type 2 remains an enigma although it might be the result of a
change in composition of the PAH family.  The larger FWHM of our class
B$_{3.3}$ likely originates from the average (through a large
aperture) of Type 2 and Type 1 / class A$_{3.3}$ profiles with a
dominating Type 2 contribution.

\subsection{The different behaviour of the CH modes versus the CC modes}

Perusing the spectra, it is striking that the variations in the 6--9
$\mu$m region are much more pronounced than those of the 3.3 or 11.2
$\mu$m features (cf., Fig~\ref{chvscc_fig:var}; Table
\ref{chvscc_tab:classProp}).  This implies that variations in the
features assigned to CC modes are significantly larger than those
assigned to CH modes. Here we discuss what could be the root of this
dichotomy.

\subsubsection{Charge state}

Experimentally and quantum chemically, it is well established that the
6--9 $\mu$m CC modes are dominated by cations while the 3.3 $\mu$m CH
stretching mode can be due to neutrals and anions
\citep[Fig.\ref{lab33} Sect.\ref{chvscc_spectroscopy_ch}, ][Mattioda
et~al., in preparation]{Szczepanski:nature:93,
Langhoff:neutionanion:96, Hudgins:tracesionezedpahs:99,
Kim:gasphasepyrenecation:01}.  There is some observational support
that this dichotomy between the CC and CH stretching modes, also
extends to the 11.2 $\mu$m CH out-of-plane bending mode (HVP01).
Specifically, the strength of the astronomical 11.2 $\mu$m mode
correlates well with the 3.3 $\mu$m CH stretching band and not with
the 6.2 $\mu$m CC stretching band.
The difference in spectral variations between CH and CC modes may
merely reflect that they probe different parts of the interstellar PAH
population.  For example, the chemical modification process might be
mainly operative on cations and not on neutrals.  In that respect,
because ionisation and neutralisation of interstellar PAHs occurs on a
very rapid timescale, this would imply a similar fast chemical
exchange. Such a fast process would also allow for spatial variations
within a source. It should be noted that it is unlikely that this
rapid behaviour is connected with the incorporation on N in PAH rings.
However, functionalized PAHs, PAH clusters or PAH-metal complexes may
be formed and broken on a similar timescale
\citep{VanKerckhoven:phd:02}.

\subsubsection{Heteroatom substitution and complexes}
Substitution of nitrogen in the ring can explain variation in the 6.2
$\mu$m band position. However, it does not influence the peak position
of the CH modes nor does it influence the peak positions of the 7.7
\mum\, complex in a systematic way. However, as mentioned above, it is
unlikely that N-incorporation deep within a PAH can occur on such a
rapid timescale as to explain the spatial variations within a
source. Complexes of PAHs with each other or with metal atoms might
also cause the spectral variation in the 6--9 \mum\, range. However,
the influence of complexation on the CH modes still has to be
investigated in the laboratory.

\subsubsection{Size}
The very small variation of the 3.3 \mum\, feature with respect to the
CC modes and to the CH out-of-plane bending modes, is likely also
related to distinct {\it subsets} of emitting PAH molecules. Indeed,
the 3.3 \mum\, feature originates from the smallest members of the
emitting PAH population while the PAH features in the 6--12 \mum\,
region originates from somewhat larger PAHs in the population
\citep{Schutte:model:93}.  This would indicate that photo-chemistry is
not the driving force behind the observed spectral variation since
smaller PAHs will be much more susceptible to unimolecular
dissociation following photon absorption and hence larger variations
would be expected for the 3.3 \mum\, feature than for higher
wavelength features, in contradiction to what is observed. In
addition, this might suggest that while the larger PAHs are modified,
the smaller ones are destroyed. This would leave only the most stable
small PAHs to emit the 3.3 \mum\, feature, while a mixture of modified
larger PAHs survive to give rise to the higher wavelength features.

\subsubsection{Molecular edge structure}
As already mentioned, any molecular structure variations, which lead
to shifts in the 6--9 $\mu$m CC bands, may be inherently accompanied
by modification of the CH peripheral structure which turn CH solo's
into duo's or trio's.

\bigskip 
To summarise, several parameters can cause the difference in the
amount of variation between the CH and CC modes. However, further
laboratory studies are required to determine which of these
parameters(s) controls the spectral characteristics of PAHs in space.
Based on the rough correlation of most of the band profiles one can
stipulate that the appearance of the UIR emission features is
determined by a few physical parameters alone. However, there are
several sources which do not follow these correlations. This implies
that the behavior cannot be understood in terms of a single
parameter.

\section{Conclusions}\label{chvscc_Concl}

Based on ISO-SWS observations, we have studied the profiles of the 3.3
and 11.2 \mum\, PAH features of a wide variety of sources and found
that clear variations are present. Both features shift in peak
position and show different profiles from source to source. Comparing
the 3.3, 6--9 and 11.2 $\mu$m classifications, we recognise a
correlation between them. In general, an $A$ classification in 6--9
$\mu$m region also implies an $A_{3.3}$ and $A_{11.2}$. However, this
seems not true for the $B$ and $C$ classifications; $B_{3.3}$ and
$B_{11.2}$ do not necessarily correlate with each other or with
$B_{6-9}$ or $C_{6-9}$. In addition, these variations depend on the
type of object considered. Apparently, a circumstellar environment
where dust formation is on-going or a protostellar/protoplanetary
environment is an essential but not sufficient condition to have
peculiar 3.3 or 11.2 \mum\, profiles.  Noteworthy is the fact that the
three galaxies in our sample show the same profile as the HII regions,
except for the 11.2 \mum\, profile which ressembles that of some
evolved stars.

Spatial variations of the profiles within a source indicate that
specific profiles are not unique to certain object types and that the
observed profile depends on the aperture. Nevertheless, the average
(or predominant) profile (as observed with ISO-SWS which has an
aperture of $14''$x$20''$) present within a source does depend on
object type.

However, the most striking aspect of the features in the 3--12 \mum\,
regions is the pronounced contrast in the profile variations between
the CH modes and the CC modes.  Specifically, the peak position in
wavenumber space varies by about 20, 80 and 15 cm$^{-1}$ for the 6.2,
7.7 and 8.6 $\mu$m features, respectively, and only by about 6.5 and
4.0 cm$^{-1}$ for the features at 3.3 and 11.2 $\mu$m, respectively.

We summarise existing laboratory data and theoretical calculations of
the modes emitting in the 3--12 \mum\, region of PAH molecules. In
contrast to the 6--9 \mum\, region which arise from PAH cations, the
3.3 \mum\, band appears to originate mainly in neutral and/or
negatively charged PAHs.

We attribute the variations in peak position of the 3.3 and 11.2
\mum\, feature to the presence of a mixture of PAH molecules, whose
composition changes. The variations in FWHM of the 3.3 \mum\, feature
remains an enigma -- although might result from the change in composition
of the PAH family -- while those of the 11.2 \mum\, can be explained by
anharmonicity and molecular structure.  The possible origin of the
observed contrast in profile variations between the CH modes and the
CC modes is highlighted.

The overall good agreement that can be achieved between the IS spectra
and the spectral characeristics of free PAHs strongly supports the PAH
model and allows these spectra to become a probe of local
conditions. However, it is clear that some features of the major band
profiles and the minor bands cannot be accounted for by "classical"
PAHs. Other PAH related species such as PAH clusters, heteroatom PAHs
and PAH-metal complexes should also be considered.

\acknowledgements EP acknowledges the support from an NWO program
  subsidy (grant number 783-70-000) and the National Research
  Council. CVK is a Research Assistant of the Fund for Scientific
  Research, Flanders.  The laboratory work was supported by NASA's
  Laboratory Astrophysics Program (grant number 344-02-04-02). This
  work is based on observations made with ISO, an ESA project with
  instruments funded by ESA member states (especially the PI
  countries: France, Germany, the Netherlands, and the United Kingdom)
  and with the participation of ISAS and NASA. IA$^3$ is a joint
  development of the SWS consortium. Contributing institutes are SRON,
  MPE, KUL and the ESA Astrophysics Division.


\begin{thebibliography}{59}
\expandafter\ifx\csname natexlab\endcsname\relax\def\natexlab#1{#1}\fi

\bibitem[{{Allamandola} {et~al.}(1999){Allamandola}, {Hudgins}, \&
  {Sandford}}]{Allamandola:modelobs:99}
{Allamandola}, L.~J., {Hudgins}, D.~M., \& {Sandford}, S.~A. 1999, \apjl, 511,
  L115

\bibitem[{{Allamandola} {et~al.}(1985){Allamandola}, {Tielens}, \&
  {Barker}}]{Allamandola:autoexhaust:85}
{Allamandola}, L.~J., {Tielens}, A.~G.~G.~M., \& {Barker}, J.~R. 1985, \apjl,
  290, L25

\bibitem[{{Allamandola} {et~al.}(1989){Allamandola}, {Tielens}, \&
  {Barker}}]{Allamandola:rev:89}
{Allamandola}, L.~J., {Tielens}, A.~G.~G.~M., \& {Barker}, J.~R. 1989, \apjs,
  71, 733

\bibitem[{{Bakes} \& {Tielens}(1994)}]{Bakes:photoelec:94}
{Bakes}, E.~L.~O. \& {Tielens}, A.~G.~G.~M. 1994, \apj, 427, 822

\bibitem[{{Bakes} {et~al.}(2001){Bakes}, {Tielens}, \&
  {Bauschlicher}}]{Bakes:modelI:01}
{Bakes}, E.~L.~O., {Tielens}, A.~G.~G.~M., \& {Bauschlicher}, C.~W. 2001, \apj,
  556, 501

\bibitem[{{Barker} {et~al.}(1987){Barker}, {Allamandola}, \&
  {Tielens}}]{Barker:anharm:87}
{Barker}, J.~R., {Allamandola}, L.~J., \& {Tielens}, A.~G.~G.~M. 1987, \apjl,
  315, L61

\bibitem[{{Bellamy}(1958)}]{Bellamy:58}
{Bellamy}, L. 1958, The infra-red spectra of complex molecules, 2nd ed. (New
  York: John Wiley {\&} Sons, Inc.)

\bibitem[{{Boulanger} {et~al.}(1998){Boulanger}, {Boisssel}, {Cesarsky}, \&
  {Ryter}}]{Boulanger:lorentz:98}
{Boulanger}, F., {Boisssel}, P., {Cesarsky}, D., \& {Ryter}, C. 1998, \aap,
  339, 194

\bibitem[{{Bregman} \& {Temi}(2004)}]{Bregman:04}
{Bregman}, J. \& {Temi}, P. 2004, \aap, in press

\bibitem[{{Brenner} \& {Barker}(1992)}]{Brenner:benz+naph:92}
{Brenner}, J. \& {Barker}, J.~R. 1992, \apjl, 388, L39

\bibitem[{{Cohen} {et~al.}(1986){Cohen}, {Allamandola}, {Tielens}, {Bregman},
  {Simpson}, {Witteborn}, {Wooden}, \& {Rank}}]{Cohen:co:86}
{Cohen}, M., {Allamandola}, L., {Tielens}, A.~G.~G.~M., {et~al.} 1986, \apj,
  302, 737

\bibitem[{{Cohen} {et~al.}(1985){Cohen}, {Tielens}, \&
  {Allamandola}}]{Cohen:85}
{Cohen}, M., {Tielens}, A.~G.~G.~M., \& {Allamandola}, L.~J. 1985, \apjl, 299,
  L93

\bibitem[{{Colangeli} {et~al.}(1992){Colangeli}, {Mennella}, \&
  {Bussoletti}}]{Colangeli:T:92}
{Colangeli}, L., {Mennella}, V., \& {Bussoletti}, E. 1992, \apj, 385, 577

\bibitem[{{Cook} \& {Saykally}(1998)}]{Cook:excitedpahs:98}
{Cook}, D.~J. \& {Saykally}, R.~J. 1998, \apj, 493, 793

\bibitem[{{Cook} {et~al.}(1998){Cook}, {Schlemmer}, {Balucani}, {Wagner},
  {Harrison}, {Steiner}, \& {Saykally}}]{Cook:uvlaserexpahs:98}
{Cook}, D.~J., {Schlemmer}, S., {Balucani}, N., {et~al.} 1998, \jpca, 102, 1465

\bibitem[{{de Graauw} {et~al.}(1996){de Graauw}, {Haser}, {Beintema},
  {Roelfsema}, {van Agthoven}, {Barl}, {Bauer}, {Bekenkamp}, {Boonstra},
  {Boxhoorn}, {Cote}, {de Groene}, {van Dijkhuizen}, {Drapatz}, {Evers},
  {Feuchtgruber}, {Frericks}, {Genzel}, {Haerendel}, {Heras}, {van der Hucht},
  {van der Hulst}, {Huygen}, {Jacobs}, {Jakob}, {Kamperman}, {Katterloher},
  {Kester}, {Kunze}, {Kussendrager}, {Lahuis}, {Lamers}, {Leech}, {van der
  Lei}, {van der Linden}, {Luinge}, {Lutz}, {Melzner}, {Morris}, {van Nguyen},
  {Ploeger}, {Price}, {Salama}, {Schaeidt}, {Sijm}, {Smoorenburg}, {Spakman},
  {Spoon}, {Steinmayer}, {Stoecker}, {Valentijn}, {Vandenbussche}, {Visser},
  {Waelkens}, {Waters}, {Wensink}, {Wesselius}, {Wiezorrek}, {Wieprecht},
  {Wijnbergen}, {Wildeman}, \& {Young}}]{deGraauw:sws:96}
{de Graauw}, T., {Haser}, L.~N., {Beintema}, D.~A., {et~al.} 1996, \aap, 315,
  L49

\bibitem[{{Flickinger} {et~al.}(1991){Flickinger}, {Wdowiak}, \&
  {Gomez}}]{Flickinger:91}
{Flickinger}, G.~C., {Wdowiak}, T.~J., \& {Gomez}, P.~L. 1991, \apjl, 380, L43

\bibitem[{{Geballe} {et~al.}(1985){Geballe}, {Lacy}, {Persson}, {McGregor}, \&
  {Soifer}}]{Geballe:85}
{Geballe}, T.~R., {Lacy}, J.~H., {Persson}, S.~E., {McGregor}, P.~J., \&
  {Soifer}, B.~T. 1985, \apj, 292, 500

\bibitem[{{Gillett} {et~al.}(1973){Gillett}, {Forrest}, \&
  {Merrill}}]{Gillett:73}
{Gillett}, F.~C., {Forrest}, W.~J., \& {Merrill}, K.~M. 1973, \apj, 183, 87

\bibitem[{{Goto} {et~al.}(2003){Goto}, {Gaessler}, {Hayano}, {Iye}, {Kamata},
  {Kanzawa}, {Kobayashi}, {Minowa}, {Saint-Jacques}, {Takami}, {Takato}, \&
  {Terada}}]{Goto:03}
{Goto}, M., {Gaessler}, W., {Hayano}, Y., {et~al.} 2003, \apj, 589, 419

\bibitem[{{Hony} {et~al.}(2001){Hony}, {Van Kerckhoven}, {Peeters}, {Tielens},
  {Hudgins}, \& {Allamandola}}]{Hony:oops:01}
{Hony}, S., {Van Kerckhoven}, C., {Peeters}, E., {et~al.} 2001, \aap, 370, 1030

\bibitem[{{Hudgins} \& {Allamandola}(2004)}]{Hudgins:colorado:04}
{Hudgins}, D.~M. \& {Allamandola}. 2004, in Astrophysics of Dust,
eds. {Witt}, A.~N. , {Clayton}, G.~C. , \& {Draine}, B.~T.,
Astronomical Society of the Pacific, in press

\bibitem[{{Hudgins} \& {Allamandola}(1999)}]{Hudgins:tracesionezedpahs:99}
{Hudgins}, D.~M. \& {Allamandola}, L.~J. 1999, \apjl, 516, L41

\bibitem[{{Hudgins} {et~al.}(1994){Hudgins}, {Sandford}, \&
  {Allamandola}}]{Hudgins:cationsI:94}
{Hudgins}, D.~M., {Sandford}, S.~A., \& {Allamandola}, L.~J. 1994, J. Phys.
  Chem., 98, 4243

\bibitem[{{Jaschek} {et~al.}(1964){Jaschek}, {Conde}, \& {de
  Sierra}}]{Jaschek:64}
{Jaschek}, C., {Conde}, H., \& {de Sierra}, A.~C. 1964, Observatory
  Astronomical La Plata Series Astronomies, 28, 1

\bibitem[{{Joblin} {et~al.}(2000){Joblin}, {Abergel}, {Bregman},
  {D'Hendecourt}, {Heras}, {Jourdain de Muizon}, {Pech}, \&
  {Tielens}}]{Joblin:isobeyondthepeaks:00}
{Joblin}, C., {Abergel}, A., {Bregman}, J., {et~al.} 2000, ISO beyond the
  peaks: The 2nd ISO workshop on analytical spectroscopy.~Eds.~A.~Salama,
  M.F.Kessler, K.~Leech \& B.~Schulz.~ESA-SP, 456, 49

\bibitem[{{Joblin} {et~al.}(1995){Joblin}, {Boissel}, {Leger}, {D'Hendecourt},
  \& {Defourneau}}]{Joblin:T:95}
{Joblin}, C., {Boissel}, P., {Leger}, A., {D'Hendecourt}, L., \& {Defourneau},
  D. 1995, \aap, 299, 835

\bibitem[{{Kerr} {et~al.}(1999){Kerr}, {Hurst}, {Miles}, \&
  {Sarre}}]{Kerr:rr:99}
{Kerr}, T.~H., {Hurst}, M.~E., {Miles}, J.~R., \& {Sarre}, P.~J. 1999, \mnras,
  303, 446

\bibitem[{{Kessler} {et~al.}(1996){Kessler}, {Steinz}, {Anderegg}, {Clavel},
  {Drechsel}, {Estaria}, {Faelker}, {Riedinger}, {Robson}, {Taylor}, \&
  {Ximenez de Ferran}}]{Kessler:iso:96}
{Kessler}, M.~F., {Steinz}, J.~A., {Anderegg}, M.~E., {et~al.} 1996, \aap, 315,
  L27

\bibitem[{{Kim} {et~al.}(2001){Kim}, {Wagner}, \&
  {Saykally}}]{Kim:gasphasepyrenecation:01}
{Kim}, H.~S., {Wagner}, D.~R., \& {Saykally}, R.~J. 2001, \prl, 86, 5691

\bibitem[{{Langhoff}(1996)}]{Langhoff:neutionanion:96}
{Langhoff}, S.~R. 1996, \jpc, 100, 2819

\bibitem[{{L\'{e}ger} {et~al.}(1989){L\'{e}ger}, {D'Hendecourt}, {Boissel}, \&
  {Desert}}]{Leger:photothermodisoc:89}
{L\'{e}ger}, A., {D'Hendecourt}, L., {Boissel}, P., \& {Desert}, F.~X. 1989,
  \aap, 213, 351

\bibitem[{{Madden} {et~al.}(2003){Madden}, {Galliano}, \&
  {Jones}}]{Madden:colorado:03}
{Madden}, S., {Galliano}, F., \& {Jones}, A. 2003, in Astrophysics of Dust,
  Colorado, 2003. Ed. by A.~N. Witt.

\bibitem[{{Maillard} {et~al.}(1999){Maillard}, {Joblin}, {Mitchell}, {Vauglin},
  \& {Cox}}]{Maillard:99}
{Maillard}, J.~P., {Joblin}, C., {Mitchell}, G.~F., {Vauglin}, I., \& {Cox}, P.
  1999, The Universe as Seen by ISO, ESA SP, 427, 707

\bibitem[{{Malfait} {et~al.}(1998){Malfait}, {Waelkens}, {Waters},
  {Vandenbussche}, {Huygen}, \& {de Graauw}}]{Malfait:crystsil}
{Malfait}, K., {Waelkens}, C., {Waters}, L.~B.~F.~M., {et~al.} 1998, \aap, 332,
  L25

\bibitem[{{Mattioda} {et~al.}(2003){Mattioda}, {Hudgins}, {Bauschlicher},
  {Rosi}, \& {Allamandola}}]{Mattioda:npahs:03}
{Mattioda}, A.~L., {Hudgins}, D.~M., {Bauschlicher}, C.~W., {Rosi}, M., \&
  {Allamandola}, L.~J. 2003, \jpc, 107, 1486

\bibitem[{{Miyata} {et~al.}(2004){Miyata}, {Kataza},{Okamoto}, {Onaka}, {Sako}, {Honda}, {Yamashita}, \& {Murakawa}}]{Miyata:rr:04}
{Miyata}, T., {Kataza}, H., {Okamoto}, Y.~K., {et~al.} 2004, \aap, 415, 179

\bibitem[{{Molster} {et~al.}(2002){Molster}, {Waters}, {Tielens}, \&
  {Barlow}}]{Molster:crystsilsample:02}
{Molster}, F.~J., {Waters}, L.~B.~F.~M., {Tielens}, A.~G.~G.~M., \& {Barlow},
  M.~J. 2002, \aap, 382, 184

\bibitem[{{Oomens} {et~al.}(2003){Oomens}, {Tielens}, {Sartakov}, {von Helden},
  \& {Meijer}}]{Oomens:03}
{Oomens}, J., {Tielens}, A.~G.~G.~M., {Sartakov}, B.~G., {von Helden}, G., \&
  {Meijer}, G. 2003, \apj, 591, 968

\bibitem[{{Pech} {et~al.}(2002){Pech}, {Joblin}, \& {Boissel}}]{Pech:prof:01}
{Pech}, C., {Joblin}, C., \& {Boissel}, P. 2002, \aap, 388, 639

\bibitem[{{Peeters} {et~al.}(2004{\natexlab{a}}){Peeters},
  {Allamandola}, {Hudgins}, {Hony}, \& {Tielens}}]{Peeters:review:04} {Peeters},
  E., {Allamandola}, L.~J., {Hudgins}, D.~M., {Hony}, S., \& {Tielens}, A.~G.~G.~M.
  2004{\natexlab{a}}, in Astrophysics of Dust, eds. {Witt}, A.~N. ,
  {Clayton}, G.~C. , \& {Draine}, B.~T., Astronomical Society of the
  Pacific, in press

\bibitem[{{Peeters} {et~al.}(2002{\natexlab{a}}){Peeters}, {Hony}, {Van
  Kerckhoven}, {Tielens}, {Allamandola}, {Hudgins}, \&
  {Bauschlicher}}]{Peeters:prof6:02}
{Peeters}, E., {Hony}, S., {Van Kerckhoven}, C., {et~al.} 2002{\natexlab{a}},
  \aap, 390, 1089

\bibitem[{{Peeters} {et~al.}(2002{\natexlab{b}}){Peeters},
  {Mart\'{\i}n-Hern\'{a}ndez}, {Damour}, {Cox}, {Roelfsema}, {Baluteau},
  {Tielens}, {Churchwell}, {Kessler}, {Mathis}, {Morisset}, \&
  {Schaerer}}]{Peeters:cataloog:01}
{Peeters}, E., {Mart\'{\i}n-Hern\'{a}ndez}, N.~L., {Damour}, F., {et~al.}
  2002{\natexlab{b}}, \aap, 381, 571

\bibitem[{{Peeters} {et~al.}(2004{\natexlab{b}}){Peeters}, {Spoon}, \&
  {Tielens}}]{Peeters:pahtracer:04}
{Peeters}, E., {Spoon}, H.~W.~W., \& {Tielens}, A.~G.~G.~M. 2004{\natexlab{b}},
  \apj, submitted

\bibitem[{{Peeters} {et~al.}(2004{\natexlab{c}}){Peeters}, {Tielens},
  {Boogert}, {Hayward}, \& {Allamandola}}]{Peeters:sc18434:04}
{Peeters}, E., {Tielens}, A.~G.~G.~M., {Boogert}, A.~C.~A., {Hayward}, T.~L.,
  \& {Allamandola}, L.~J. 2004{\natexlab{c}}, \aap, submitted

\bibitem[{{Puget} \& {L\'{e}ger}(1989)}]{Puget:revpah:89}
{Puget}, J.~L. \& {L\'{e}ger}, A. 1989, \araa, 27, 161

\bibitem[{{Roche} {et~al.}(1989){Roche}, {Aitken}, \& {Smith}}]{Roche:orion:89}
{Roche}, P.~F., {Aitken}, D.~K., \& {Smith}, C.~H. 1989, \mnras, 236, 485

\bibitem[{{Schutte} {et~al.}(1993){Schutte}, {Tielens}, \&
  {Allamandola}}]{Schutte:model:93}
{Schutte}, W.~A., {Tielens}, A.~G.~G.~M., \& {Allamandola}, L.~J. 1993, \apj,
  415, 397

\bibitem[{{Song} {et~al.}(2003{\natexlab{a}}){Song}, {Kerr}, {McCombie}, \&
  {Sarre}}]{Song:33:03}
{Song}, I., {Kerr}, T., {McCombie}, J., \& {Sarre}, P. 2003{\natexlab{a}},
  \mnras, 346, L1

\bibitem[{{Song} {et~al.}(2003{\natexlab{b}}){Song}, {McCombie},
  {Kerr}, {Couch}, \& {Sarre}}]{Song:colorado:03} {Song}, I.,
  {McCombie}, J., {Kerr}, T., {Couch}, P., \& {Sarre}, P.
  2003{\natexlab{b}}, in Astrophysics of Dust, Colorado, 2003. Ed. by
  A.~N. Witt.

\bibitem[{{Szczepanski} \& {Vala}(1993)}]{Szczepanski:nature:93}
{Szczepanski}, J. \& {Vala}, M. 1993, \nat, 363, 699

\bibitem[{{Tielens} {et~al.}(1999){Tielens}, {Hony}, {Van Kerckhoven}, \&
  {Peeters}}]{Tielens:parijs:99}
{Tielens}, A.~G.~G.~M., {Hony}, S., {Van Kerckhoven}, C., \& {Peeters}, E.
  1999, in ESA SP-427: The Universe as Seen by ISO, Vol. 427, 579

\bibitem[{{Tokunaga} {et~al.}(1991){Tokunaga}, {Sellgren}, {Smith}, {Nagata},
  {Sakata}, \& {Nakada}}]{Tokunaga:33prof:91}
{Tokunaga}, A.~T., {Sellgren}, K., {Smith}, R.~G., {et~al.} 1991, \apj, 380,
  452

\bibitem[{{Van Kerckhoven}(2002)}]{VanKerckhoven:phd:02}
{Van Kerckhoven}, C. 2002, PhD thesis, Katholieke Universiteit Leuven (Belgium)

\bibitem[{{Vermeij} {et~al.}(2002){Vermeij}, {Peeters}, {Tielens}, \& {van der
  Hulst}}]{Vermeij:pahs:01}
{Vermeij}, R., {Peeters}, E., {Tielens}, A.~G.~G.~M., \& {van der Hulst}, J.~M.
  2002, \aap, 382, 1042

\bibitem[{{Verstraete} {et~al.}(2001){Verstraete}, {Pech}, {Moutou},
  {Sellgren}, {Wright}, {Giard}, {L{\' e}ger}, {Timmermann}, \&
  {Drapatz}}]{Verstraete:prof:01}
{Verstraete}, L., {Pech}, C., {Moutou}, C., {et~al.} 2001, \aap, 372, 981

\bibitem[{{Verstraete} {et~al.}(1996){Verstraete}, {Puget}, {Falgarone},
  {Drapatz}, {Wright}, \& {Timmermann}}]{Verstraete:m17:96}
{Verstraete}, L., {Puget}, J.~L., {Falgarone}, E., {et~al.} 1996, \aap, 315,
  L337

\bibitem[{{Wada} {et~al.}(2003){Wada}, {Onaka}, {Yamamura}, {Murata}, \& {Tokunaga}}]{Wada:13c:03}
  {Wada}, S., {Onaka}, T., {Yamamura}, I., {Murata}, Y., {Tokunaga}, A.~T. 2003, \aap, 407, 551

\bibitem[{{Wagner} {et~al.}(2000){Wagner}, {Kim}, \& {Saykally}}]{Wagner:2000}
{Wagner}, D.~R., {Kim}, H., \& {Saykally}, R.~J. 2000, \apj, 545, 854


\bibitem[{{Witteborn} {et~al.}(1989){Witteborn}, {Sandford}, {Bregman},
  {Allamandola}, {Cohen}, {Wooden}, \& {Graps}}]{Witteborn:89}
{Witteborn}, F.~C., {Sandford}, S.~A., {Bregman}, J.~D., {et~al.} 1989, \apj,
  341, 270

\end{thebibliography}
\end{document}